\documentclass[english,prb,twocolumn,notitlepage,superscriptaddress,showpacs]{revtex4-1}

\usepackage{graphicx}
\usepackage{color}
\usepackage{amssymb}
\usepackage{amsmath}
\usepackage{setspace}
\usepackage{babel}
\usepackage[utf8x]{inputenc}

\begin{document}

\title{Effective Hamiltonian for the hybrid double quantum dot qubit}
\author{E. Ferraro}
\email{elena.ferraro@mdm.imm.cnr.it}
\author{M. De Michielis}
\affiliation{Laboratorio MDM, IMM-CNR, Via Olivetti 2, I-20864 Agrate Brianza, Italy}
\author{G. Mazzeo}
\affiliation{Laboratorio MDM, IMM-CNR, Via Olivetti 2, I-20864 Agrate Brianza, Italy}
\author{M. Fanciulli}
\affiliation{Laboratorio MDM, IMM-CNR, Via Olivetti 2, I-20864 Agrate Brianza, Italy}
\affiliation{Dipartimento di Scienza dei Materiali, University of Milano Bicocca, Via R. Cozzi, 53, 20126 Milano, Italy}
\author{E. Prati}
\affiliation{Laboratorio MDM, IMM-CNR, Via Olivetti 2, I-20864 Agrate Brianza, Italy}

\begin{abstract}
Quantum dot hybrid qubits formed from three electrons in double quantum dots represent a promising compromise between high speed and simple fabrication for solid state implementations of single qubit and two qubits quantum logic ports. We derive the Schrieffer-Wolff effective Hamiltonian that describes in a simple and intuitive way the qubit by combining a Hubbard-like model with a projector operator method. As a result, the Hubbard-like Hamiltonian is transformed in an equivalent expression in terms of the exchange coupling interactions between pairs of electrons. The effective Hamiltonian is exploited to derive the dynamical behaviour of the system and its eigenstates on the Bloch sphere to generate qubits operation for quantum logic ports. A realistic implementation in silicon and the coupling of the qubit with a detector are discussed.
\end{abstract}

\maketitle
\section{Introduction}
Dynamics of the quantum dot hybrid qubit is obtained from the Schrieffer-Wolff transformed effective Hamiltonian in terms of effective exchange coupling interaction between pairs of electrons.

Spin dynamics in quantum dot have attracted wide attention in the scientific community both from experimental\cite{exp1,exp2,exp3,exp4,exp5,exp6} and theoretical\cite{theo1,theo2,theo3} point of view because of their long coherence times and potential scaling \cite{spin1,spin2,spin3}.

Several architectures have been proposed based on single \cite{spin1}, double \cite{spin3,st1,st2} and triple \cite{spin2} quantum dot, later implemented in GaAs \cite{as1,as2,as3,as4}, Si \cite{si1,si2,si3,si4} and InSb \cite{in} nanostructures. In view of creating an architecture capable to assure the best compromise among fabrication, tunability, fast gate operations of one and two qubits, manipulability and scalability, hybrid qubits have been recently proposed \cite{Shi}. They consist of a double quantum dot with three electrons distributed during the operations between the two quantum dots, with at least one electron in each. The interest raised by this architecture is due to the possibility to obtain gate operations entirely implemented with purely electrical manipulations. This property enables much faster gate operations than using ac magnetic fields, inhomogeneous dc magnetic fields or mechanisms based on spin-orbit coupling. The Heisenberg interaction which is the dominant mechanism of interaction between adjacent spins suffice for all the one- and two-qubits operations, removing the need of an inhomogeneous field which is required for instance in single-triplet qubits \cite{spin3,st1,st2}. Not so surprisingly, such an architecture grants the same symmetries in spin space as the triple dot qubit proposed in Ref.\cite{spin2}, but it is simpler and more compact to fabricate, as it requires only two dots instead of three.

Starting from the Hubbard-like model, we derive a general effective Hamiltonian for the hybrid qubit in terms of the spin operators of the three electrons and of their interactions, by defining a suitable projection operator that follows the method of Schrieffer and Wolff \cite{sw}. The method here adopted enables us to obtain analytically the coupling constants between pairs of electrons under lesser restrictive conditions than those obtained in Ref.\cite{Shi}, preserving an explicit dependence of all the parameters involved. The system is irrespective to the host material, whose properties are absorbed in the coupling coefficients. Next, dynamics generated by the interaction terms is expressed. The Heisenberg interaction should permit very fast gate operation, however by itself does not provide a universal gate. It cannot generate any arbitrary unitary transformation on a collection of spin-1/2 qubits as it allows only rotations among states with the same quantum numbers. However, defining opportunely coded qubit states the Heisenberg interaction alone is universal.

The paper is organized as follows. Section II is devoted to the derivation of the effective Hamiltonian for a single hybrid qubit and its dynamical behaviour, once realistic initial conditions are provided and a realistic implementation in silicon is studied in stationary conditions. In Section III the pictorial representation of the eigenvectors in the Bloch sphere analyzing two limiting cases of interest, corresponding to the situation in which two electrons are confined in the right or in the left dot, is presented. 

\section{Effective Hamiltonian for the hybrid qubit}
\begin{figure}[h]
\begin{center}
\includegraphics[width=0.2\textwidth]{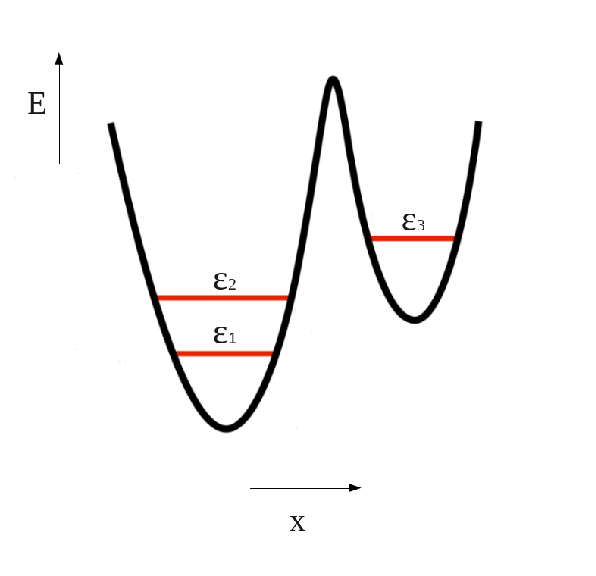}
\end{center}
\caption{Schematic energy profile and single electron energy levels in the hybrid spin qubit. Each level has a two-fold spin degeneracy.}\label{hybrid}
\end{figure}
This Section describes the derivation of an effective Hamiltonian that accounts for the quantum behaviour of the hybrid qubit in the regime of low energy excitations. Next, the dynamical behaviour in the logical subspace is analyzed. The hybrid qubit, whose energy profile is shown schematically in Fig. \ref{hybrid}, is composed of a double quantum dot with two electrons in the left dot and one electron in the right dot. Levels denoted by $\varepsilon_1$ and $\varepsilon_2$ refer to the lowest single particle energies of the doubly occupied dot in the left, $\varepsilon_3$ is the lowest single particle energy of the single occupied dot on the right. Each level has a two-fold degeneracy.

\subsection{Effective Hamiltonian from the projector method}
The Hubbard-like Hamiltonian \cite{hubbard} model of the double dot hybrid qubit is written in terms of the creation and annihilation fermionic operators of the three spins, which are $c_k^{\dagger}$ and $c_k$ respectively, and consists at first order of four contributions, given by
\begin{equation}\label{hubbard}
H=H_e+H_t+H_U+H_J
\end{equation}
with
\begin{align}\label{hubbard2}
H_e&=\sum_{k=1,\sigma}^{3}\varepsilon_kc^{\dagger}_{k\sigma}c_{k\sigma}\nonumber\\
H_t&=t_{13}\sum_{\sigma}(c^{\dagger}_{1\sigma}c_{3\sigma}+h.c.)+t_{23}\sum_{\sigma}(c^{\dagger}_{2\sigma}c_{3\sigma}+h.c.)\nonumber\\
H_U&=\sum_{k=1}^3U_kn_{k\uparrow}n_{k\downarrow}+U_{12}(n_{1\uparrow}+n_{1\downarrow})(n_{2\uparrow}+n_{2\downarrow})+\nonumber\\
&+U_{13}(n_{1\uparrow}+n_{1\downarrow})(n_{3\uparrow}+n_{3\downarrow})+U_{23}(n_{2\uparrow}+n_{2\downarrow})(n_{3\uparrow}+n_{3\downarrow})\nonumber\\
H_J&=H^{13}_J+H^{23}_J+H^{12}_J
\end{align}
where
\begin{align}\label{j}
H^{ij}_J&=-J^{ij}_e(n_{i\uparrow}n_{j\uparrow}+n_{i\downarrow}n_{j\downarrow})-\left(J^{ij}_ec^{\dagger}_{i\downarrow}c^{\dagger}_{j\uparrow}c_{j\downarrow}c_{i\uparrow}+\right.\nonumber\\
&\left.+J^{ij}_pc^{\dagger}_{j\uparrow}c^{\dagger}_{j\downarrow}c_{i\uparrow}c_{i\downarrow}+\sum_{k,\sigma}J^{ij}_tn_{k\sigma}c^{\dagger}_{i\bar{\sigma}}c_{j\bar{\sigma}}+h.c.\right).
\end{align}
The first two terms $H_e$ and $H_t$ describe respectively the single electron energy level of each dot and the tunneling energy between the two dots. These two terms together constitute the single-particle part of the Hamiltonian. The last two terms $H_U$ and $H_J$ constitute instead the two-body part, i.e., the Coulomb interaction between pairs of electrons belonging to the same dot (intra-dot interaction) or to different dot (inter-dot interaction). Taking into consideration only the two-body part in correspondence to $H_U$ it was observed only an anti-ferromagnetic exchange coupling between the pair of electrons \cite{theo2}. Numerical results based on Heitler-London approximation or Hund-Mulliken molecular-orbit method have instead shown that the exchange coupling can be ferromagnetic under certain parameters conditions \cite{exchange1,exchange2}. Here, we choose to start our analysis from this exact second quantized Hamiltonian (\ref{hubbard}) to express exchange couplings between pair of electron spins. In particular the coefficients here introduced in Eq.(\ref{j}) are: the spin-exchange $J^{ij}_e$, the pair-hopping $J^{ij}_p$ and the occupation-modulated hopping terms $J^{ij}_t$. These parameters, as well as the Coulomb energies and the intra-dot bias voltage, are fixed by the geometry of the system. On the contrary, the tunneling parameters and the inter-dot bias voltage are tunable, by acting on the potential barriers.

The Hamiltonian (\ref{hubbard}) is diagonalized in the restricted Hilbert space defined by Slater determinants constructed from orthogonalized one-electron wave functions, that give an accurate low-energy spectrum. It is well known that higher-lying base states may be accounted for by perturbation theory, for example in lattice model. We shall not attempt such a reduction here for the quantum dot problem, but we merely point out its main effect would be a renormalization of the parameters involved. The high-energy states of the Hamiltonian are eliminated to yield an effective spin Hamiltonian using degenerate perturbation theory or a canonical transformation, such as the Schrieffer-Wolff transformation.

In the same spirit of Ref. \cite{theo2}, by emplyoing the projector operator method \cite{sw}, we convert the Hubbard-like Hamiltonian (\ref{hubbard}) into its effective equivalent with spin-spin interactions, i.e. exchange coupling, between pairs of electrons. We define the projector $P$ and its complementary $Q$, 
\begin{equation}
P=\Pi_{k=1}^3[n_{k\uparrow}(1-n_{k\downarrow})+n_{k\downarrow}(1-n_{k\uparrow})]\qquad Q=1-P,
\end{equation}
in such a way that the total Hilbert space of the three electrons system is divided into a subspace $P$, which consists of holes and single occupancies and the complementary subspace $Q$ with at least one double occupancy. We point out that a system with only spin degeneracy is considered. This holds for both III -- V semiconductors for which direct bandgap sussists, as well as valley degenerate indirect bandgap semiconductors for which the degeneracy is lifted in nanoscale devices \cite{marco}. 

\begin{figure}[h]
\begin{center}
\includegraphics[width=0.45\textwidth]{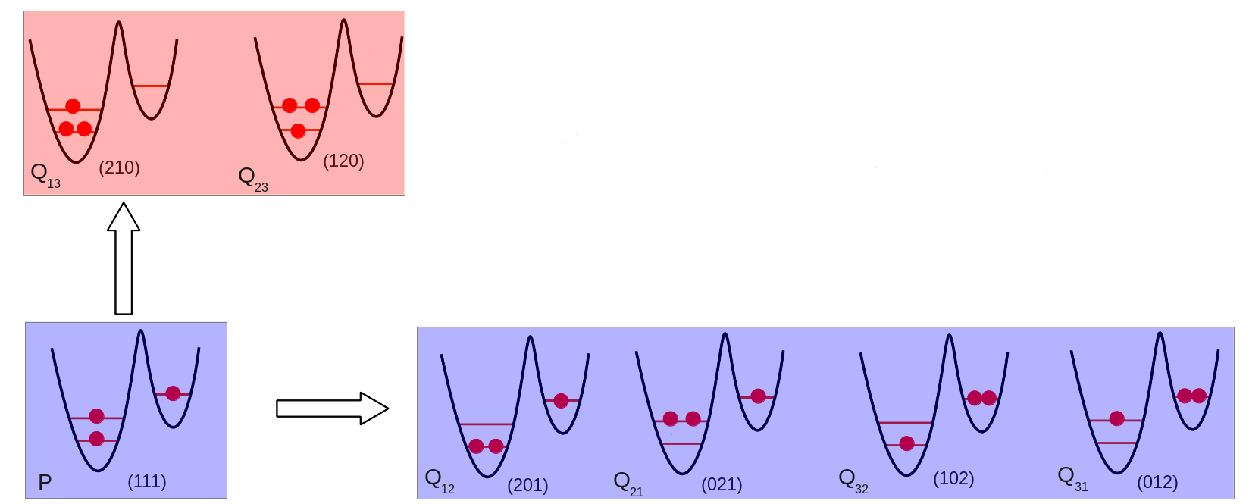}
\end{center}
\caption{Configurations in correspondence to the projectors P and $Q_{ij}$. The configuration in the blue box at the left is the one of reference and correspond to the projection P. The configurations in the blue box on the right, having the energy of the same order of magnitude of the configuration on the left, contribute in a consistent way to the effective Hamiltonian. Those in the red box in the upper part have higher energies and are neglected.}\label{inter} 
\end{figure}

It is possible to demonstrate starting from the Schr\"odinger equation \cite{theo3} that 
\begin{equation}
H^{eff}=PHP-PHQ\frac{1}{QHQ-E}QHP,
\end{equation}
describes as well the dynamics of $H$. In addition we observe that the operator $Q$ can be expanded as $Q=\sum_{i\neq j}Q_{ij}$ and concerning with the low energy excitation the term
\begin{equation}
Q_{ij}\frac{1}{QHQ-E}Q_{ij}
\end{equation}
could be replaced by $\frac{Q_{ij}}{\Delta E_{ij}}$, where $\Delta E_{ij}$ is the energy difference of the energy with one $Q_{ij}$ and the energy of $PHP$. Finally the effective Hamiltonian appears in the following form
\begin{equation}\label{eff}
H^{eff}\approx PHP-\sum_{i\neq j}\frac{1}{\Delta E_{ij}}PHQ_{ij}HP.
\end{equation}
To calculate all the contributions in Eq.(\ref{eff}), let's consider initially all the possible configurations with single and double occupancies in the three orbitals. The configurations are shown in Fig. \ref{inter} and are divided in different boxes with different colours in such a way to have at our disposal a visualization of the terms that contribute in a consistent way to the calculations, neglecting those with higher energy. 

For completeness we provide the explicit expression of the energies associated to each configuration: 
\begin{align}
E_{(ijk)}&=i\varepsilon_1+j\varepsilon_2+k\varepsilon_3+ijU_{12}+ikU_{13}+kjU_{23}+\nonumber\\
&+\delta_{i2}U_1+\delta_{j2}U_2+\delta_{k2}U_3
\end{align}
where the indices $0\le i\neq j\neq k\le2$, assuming only integer values, denote the number of electrons in each energy level. The quantities of interest, appearing in Eq.(\ref{eff}) are the differences of $E_{(ijk)}$ with the energy associated to $PHP$, corresponding to the case in which there is only one electron in each orbital, that is
\begin{equation}
E_{(111)}=\varepsilon_1+\varepsilon_2+\varepsilon_3+U_{12}+U_{13}+U_{23}.
\end{equation}
In order to separate the main contribution from perturbations at this point we evaluate the order of magnitude of the energies involved: the inter-dot energies indeed are smaller with respect to the intra-dot ones, that is 
\begin{equation}\label{eps}
\Delta\varepsilon_{12}\equiv\varepsilon_1-\varepsilon_2\ll U_{13}\sim U_{23}\ll U_1, U_2, U_3, U_{12},
\end{equation}
$\Delta\varepsilon_{13}$ and $\Delta\varepsilon_{23}$ are tunable. This condition allows us to neglect some terms with respect to others since they contribute less. In particular the neglected terms are those in correspondence to the highest energy differences (depicted in the red box in Fig. \ref{inter}) that are of the order of magnitude of $U_1$. 

Under such conditions, the contributions appearing in Eq.(\ref{eff}) return:
\begin{align}
PHP&=\sum_{k=1}^{3}\varepsilon_k+U_{12}+U_{13}+U_{23}-\frac{1}{2}(J_e^{13}+J_e^{23}+J_e^{12})+\nonumber\\
&-2J_e^{13}(\bold{S}_1\cdot\bold{S}_3+1/4)-2J_e^{23}(\bold{S}_2\cdot\bold{S}_3+1/4)+\nonumber\\
&-2J_e^{12}(\bold{S}_1\cdot\bold{S}_2+1/4)
\end{align}
\begin{align}
&PHQ_{12}HP=PHQ_{21}HP=4(J^{12}_t)^2(1/4-\bold{S}_1\cdot\bold{S}_2)\nonumber\\
&PHQ_{13}HP=PHQ_{31}HP=4(t_{13}-J^{13}_t)^2(1/4-\bold{S}_1\cdot\bold{S}_3)\nonumber\\
&PHQ_{23}HP=PHQ_{32}HP=4(t_{23}-J^{23}_t)^2(1/4-\bold{S}_2\cdot\bold{S}_3),
\end{align}
where the following identities, that directly link the fermionic operators with the spin operators, have been used :
\begin{align}
&c_{k\sigma}^{\dagger}c_{k\sigma'}=\frac{1}{2}\delta_{\sigma\sigma'}(n_{k\uparrow}+n_{k\downarrow})+\bold{S}_k\cdot\boldsymbol{\sigma}_{\sigma'\sigma}\nonumber\\
&c_{k\sigma}c_{k\sigma'}^{\dagger}=\delta_{\sigma\sigma'}\left[1-\frac{1}{2}(n_{k\uparrow}+n_{k\downarrow})\right]-\bold{S}_k\cdot\boldsymbol{\sigma}_{\sigma\sigma'},
\end{align}
where the spin operator is defined by $\bold{S}_k=\frac{1}{2}\sum_{\sigma,\sigma'}c_{k\sigma}^{\dagger}\boldsymbol{\sigma}_{\sigma\sigma'}c_{k\sigma'}$ and $\boldsymbol{\sigma}$ is the Pauli operator.
 
Collecting finally all the terms and upon neglecting the constant factors, we finally reach the closed form for the effective Hamiltonian
\begin{equation}\label{Heff}
H^{eff}\approx J_{13}\bold{S}_1\cdot\bold{S}_3+J_{23}\bold{S}_2\cdot\bold{S}_3+J_{12}\bold{S}_1\cdot\bold{S}_2
\end{equation}
where, neglecting the configurations $Q_{13}$ and $Q_{23}$, that give the highest energy differences, the effective coupling constants are given by
\begin{align}\label{coupling}
&J_{13}\equiv J_1\simeq\frac{1}{E_{(012)}-E_{(111)}}4(t_{13}-J^{13}_t)^2-2J^{13}_e\nonumber\\
&J_{23}\equiv J_2\simeq\frac{1}{E_{(102)}-E_{(111)}}4(t_{23}-J^{23}_t)^2-2J^{23}_e\nonumber\\
&J_{12}\equiv J'=\left(\frac{1}{E_{(201)}-E_{(111)}}+\frac{1}{E_{(021)}-E_{(111)}}\right)4(J^{12}_t)^2-2J^{12}_e.
\end{align}
The exchange couplings include the effects of dot tunneling, dot bias, and both on-site and off-site Coulomb interactions. Each coupling term contains the term $-2J_e^{ij}$ which is the ferromagnetic direct exchange between the two electrons from their Coulomb interactions and the anti-ferromagnetic superexchange. Consequently, the value of each $J_{ij}$ can be either positive or negative and it depends stricly on the values of the parameters entering into the expressions (\ref{coupling}). The coupling constants are tunable thanks to the control on the tunneling couplings $t_{ij}$ which can be provided by external gates and on the inter-dot bias voltage $\Delta\varepsilon_{ij}$. On the contrary, the Coulomb energy $U_i$, $U_{ij}$, $J_e^{ij}$ and $J_t^{ij}$, as well as the intra-dot bias voltage $\Delta\varepsilon_{12}$, are geometry dependent and they cannot be easily tuned. These approximations reflect the realistic conditions in which a quantum dot is operated. We point out that if condition in Eq.(\ref{eps}) is neglected, the effective couplings (\ref{coupling}) maintain the same structure and the same dependence on the physical parameters, such as the tunneling, but contain also the energy associated to the intermediate states in the red box in Fig. \ref{inter}.

In this analysis we have neglected the Overhauser field from the lattice nuclear spins. It is neglegible for fast operations or for purified silicon, for which the hyperfine coupling can be neglected with respect to the strong exchange coupling between pairs of spins. In all the other cases a more detailed analysis including the nuclear field operator in each dot is needed.

We have reached a closed form for the effective Hamiltonian describing the hybrid qubit constituted by only direct exchange interactions between pair of spins. The dependence of all the physical parameters of the dots is preserved.

\subsection{Qubit base and matrix expression of the effective Hamiltonian}
In this paragraph the effective Hamiltonian is recast in a more compact and useful form using the logical basis introduced in Ref.\cite{Shi}. With the aim of defining the logical basis for the hybrid qubit, we first enumerate the possible transitions between the three electrons spin states that can be induced by manipulations which preserve total spin angular momentum. To do this we consider the Hilbert space of three electron spins that has a dimension of $8$. The total spin eigenstates form a quadruplet with $S=3/2$ and $S_z=-3/2,-1/2,1/2,3/2$ and two doublets each with $S=1/2$ and $S_z=\pm1/2$, where the square of the total spin is $\hbar^2S(S+1)$ and the $z$-component of the total spin is $\hbar S_z$. To encode our qubit we restrict to the two-dimensional subspace of three-spin states with spin quantum numbers $S=1/2$ and $S_z=-1/2$, like in Ref.\cite{Shi}. We point out that only states with the same S and $S_z$ can be coupled by spin independent terms in the Hamiltonian. The logical basis $\{|0\rangle,|1\rangle\}$, that we are going to introduce, is constituted by singlet and triplet states of a pair of spins, for example the pair in the left dot, in combination with the angular momentum of the third spin, localized in the right dot. This means that the logical states choosen are finally expressed in this way
\begin{equation}\label{01}
|0\rangle\equiv|S\rangle|\downarrow\rangle, \qquad |1\rangle\equiv\sqrt{\frac{1}{3}}|T_0\rangle|\downarrow\rangle-\sqrt{\frac{2}{3}}|T_-\rangle|\uparrow\rangle
\end{equation}
where $|S\rangle$, $|T_0\rangle$ and $|T_-\rangle$ are respectively the singlet and triplet states, whose explicit form, in terms of the eigenstates of $\sigma_z$, is here reported for completeness
\begin{equation}
|S\rangle=\frac{|\uparrow\downarrow\rangle-|\downarrow\uparrow\rangle}{\sqrt{2}}, \quad |T_0\rangle=\frac{|\uparrow\downarrow\rangle+|\downarrow\uparrow\rangle}{\sqrt{2}}, \quad |T_-\rangle=|\downarrow\downarrow\rangle.
\end{equation}
Explicit calculations of the matrix elements of $H^{eff}$ in this basis give
\begin{align}
&\langle0|H^{eff}|0\rangle=-\frac{3}{4}J'\nonumber\\
&\langle1|H^{eff}|1\rangle=\frac{1}{4}J'-\frac{1}{2}(J_1+J_2)\nonumber\\
&\langle0|H^{eff}|1\rangle=\langle1|H^{eff}|0\rangle=-\frac{\sqrt{3}}{4}(J_1-J_2),
\end{align}
leading directly to the following matrix form
\begin{equation}\label{effmatrix}
H^{eff}= \begin{pmatrix}
-\frac{3}{4}J' & -\frac{\sqrt{3}}{4}(J_1-J_2) \\
-\frac{\sqrt{3}}{4}(J_1-J_2) & \frac{1}{4}J'-\frac{1}{2}(J_1+J_2)\end{pmatrix}.
\end{equation}

The effective Hamiltonian just derived (\ref{effmatrix}) could be successfully exploited to analyze the dynamical behaviour of the system. Eq.(\ref{effmatrix}) contains indeed all the informations on the system through the coupling constants, that when varied from the external allow operations on the qubit. Gate operations are implemented by acting on the energy splitting and by increasing the tunnel couplings, which drive transitions between qubit states. We point out that matrix elements of the Hamiltonian depend on the tunnel amplitude and are inversely proportional to the energy difference. We conclude this Section by presenting an example of the dynamical evolution of the system by starting from a specific initial condition.

We start from the initial condition in which the qubit is prepared in the state denoted by $|0\rangle$ in the logical basis. In Appendix \ref{A} the full exact dynamics is derived and the explicit probability amplitudes concerning the state of the system at the time instant $t$ are given, once the initial condition is fixed. Exploiting Eqs.(\ref{10}) we obtain the condition on the time instant under which the qubit undergoes a transition from the logical state $|0\rangle$ to $|1\rangle$
\begin{equation}\label{t}
t=\frac{2}{\sqrt{(A-B)^2+4C^2}}\arcsin\left(\pm\frac{\sqrt{(A-B)^2+4C^2}}{2C}\right),
\end{equation}
where the coefficients $A$, $B$ and $C$ are defined in Eqs.(\ref{abc}). We notice that the $\arcsin$ function is defined in the interval $[-1,1]$, as a consequence there is only one condition for the coefficients of Eq.(\ref{t}) returning real valued time, that is 
\begin{equation}\label{condition}
A\simeq B\Rightarrow t\simeq\frac{1}{C}\arcsin(\pm1).
\end{equation}
The condition for the coefficients in terms of physical parameters of the system is expressed by
\begin{equation}
3J'\simeq\sqrt{3}(J_1-J_2).
\end{equation}
This relationship gives a simple condition that links directly the coupling constants involved in the system. It is consequently possible to change suitably the tunnel barrier between the dots to guide the system toward a switch from $|0\rangle$ to $|1\rangle$.

\subsection{Silicon hybrid qubit}
In this subsection we study an example of the implementation of the hybrid qubit in silicon. Differently from bulk silicon which has a conduction band structure with three couples of energy degenerate valleys, in silicon quantum dots a sequence of energy levels is present with their valley degeneracy removed by interface delta-like potentials\cite{fri}. In the past, some of the authors already explored the creation of robust qubits in silicon and in particular it has been applied the current spin density functional theory (CS-DFT) to investigate the geometrical effects on the filling of the lowest valley states in quantum dot formed in single electron transistors (SET) \cite{marco}. The SET was modeled as a silicon nanowire on the top of a silicon dioxide slab where a tri-sided metal gate, insulated from the nanowire with a silicon dioxide layer and from the source/drain contacts with two silicon nitride spacers, electrostatically forms a quantum dot, like in Ref.\cite{prati}. By using this model, the effects of the fabrication variability of the sizes of the device on the valley filling and addition energy patterns were studied. It has been shown in Ref.\cite{marco} that variations of width and thickness of the nanowire with squared sections entail important changes in the valley filling sequences, whereas the variability of the gate length does not imply a significant modification of the filling patterns. The filling of states belonging only to one type of valley is possible for the first electrons if the silicon thickness is much smaller than the width of the nanowire. As a result, a high valley orbit energy can be obtained so that higher quantum dot energy levels related to other valleys can be neglected leaving a single fundamental valley. The silicon 2quantum dot system, studied in this section, is well described by the single valley effective Hamiltonian (19) if we assume a high valley orbit energy in both quantum dots.

\begin{figure}[t]
\includegraphics[width=0.5\textwidth]{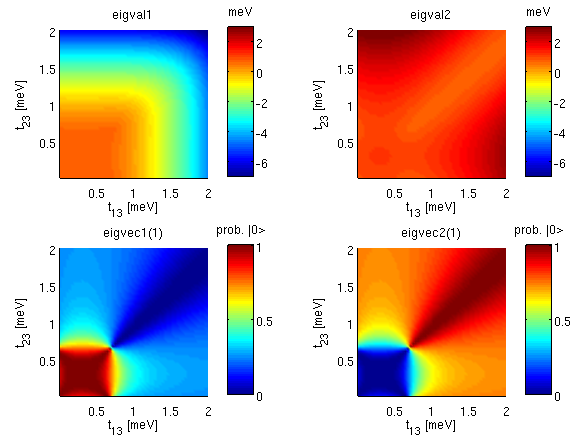}
\caption{Eigenvalues (top) and probabilities (bottom) to find the eigenvectors in $|0\rangle$ as a function of the tunneling couplings $t_{13}$ and $t_{23}$ for the first eigenstate (left) and the second one (right). The parameters used in the simulation are:
$\varepsilon_{1}$=0 meV, $\varepsilon_{2}$=0.3 meV, $\varepsilon_{3}$=0.35 meV, $U_1$=9.8 meV, $U_2$=9.8 meV, $U_3$=11 meV, $U_{13}$=1.8 meV, $U_{23}$=1.8 meV, $U_{12}$=9.8 meV, $J_{t}^{12}$=0.3 meV, $J_{t}^{13}$=0.3 meV, $J_{t}^{23}$=0.3 meV, $J_{e}^{12}$=0.5 meV, $J_{e}^{13}$=0.7 meV and $J_{e}^{23}$=0.7 meV.}\label{Fig:Eigen-t2x2} 
\end{figure}

Focusing on the stationary solutions of the Hamiltonian, in Fig. \ref{Fig:Eigen-t2x2} we report the eigenvalues and the probabilities of the eigenvectors to be found in $|0\rangle$ as a function of the tunneling couplings $t_{13}$ and $t_{23}$. For the simulations most of the parameters of the effective Hamiltonian are taken from Ref.\cite{Wang2011}, whereas $J_t^{ij}$ and $J_e^{ij}$ have been increased because they were slightly underestimated as stated in Ref.\cite{Wang2011}. The system is highly sensitive to the choice of those parameters and we set them to have a good control over the eigenstates of the system in the range of the tunnelling couplings considered. The energy gap between the eigenvalues is small for low tunneling couplings whereas it increases for higher $t_{13}$ and $t_{23}$ values. Looking at the first eigenvector, the probability to be in $|0\rangle$ is closer to 1 when $t_{12}$ and $t_{13}$ are low. Increasing $t_{12}$ and $t_{13}$ the probability to be in $|0\rangle$ instead decreases to 0 in such a way that the first eigenvector is closer to $|1\rangle$ state. The opposite behaviour is observed for the second eigenvector.

On the contrary, if exactly the same parameters values of Ref.\cite{Wang2011} are used, the tunnelling couplings are not so effective in controlling the eigenstates of the system, leaving the lowest eigenvector close to $|0\rangle$ (not shown).

The read out of the qubit state coincides with the read out of the spin state of electrons confined in the doubly occupied quantum dot. A SET can be used to electrostatically sense the spin state. The SET and the quantum dot are coupled through an energy barrier controlled electrostatically by a gate. During manipulation time interval, the electrostatic barrier between the doubly occupied quantum dot and the SET is high thus both the wavefunctions of the electrons do not interact, causing a negligible effect on the spin state of the electron couple. When read out starts, tunneling is allowed from the doubly occupied quantum dot to the SET by a reduction of the interposed electrostatic barrier. When the couple of electrons are in a singlet state the corresponding wavefunction is more confined and the tunneling rate to the SET is lower than that of the triplet state which has a broader wavefunction. When the electron tunnels, the electrostatic potential landscape changes and so does the current passing through the electrostatically coupled SET \cite{exp4,si2}. The measurement of the time interval between the read out signal and the current variation in the SET reveals the spin state of the electron couple.
A high number of electrons have to be collected in the SET to mix its collective spin state, reducing the perturbing effect of SET/quantum dot mutual exchange interactions on the spin state of the doubly occupied quantum dot. In this way, the final read out result is mostly determined by the spin state of the electron couple in the quantum dot.

\section{Qubit operations: Bloch sphere}
The objective of this Section, exploiting the effective Hamiltonian (\ref{Heff}) obtained in Section II, is to analyze how exchange interaction can drive arbitrary qubit operations. Eigenstates are here derived considering the full basis of the three electrons and represented as points on the Bloch sphere in connection with the states of the logical basis presented earlier. The single qubit operation is directly linked to the variation of the effective coupling, that appears as a rotation around a specific axis on the Bloch sphere.

Eigenstates and eigenenergies of the three electron spins in two dots are shown in Eqs.(\ref{eig}). As we have already mentioned, the eight spin eigenstates of the Hamiltonian comprise a quadruplet $|Q_{S_z}\rangle$ and high- and low-energy doublets $|\Delta_{S_z}\rangle$ and $|\Delta'_{S_z}\rangle$,
\begin{widetext}
\begin{align}\label{eig}
&|Q_{+\frac{3}{2}}\rangle=|\uparrow\uparrow\uparrow\rangle\nonumber\\
&|Q_{+\frac{1}{2}}\rangle=\frac{1}{\sqrt{3}}(|\uparrow\uparrow\downarrow\rangle+|\uparrow\downarrow\uparrow\rangle+|\downarrow\uparrow\uparrow\rangle)\nonumber\\
&|Q_{-\frac{1}{2}}\rangle=\frac{1}{\sqrt{3}}(|\downarrow\downarrow\uparrow\rangle+|\downarrow\uparrow\downarrow\rangle+|\uparrow\downarrow\downarrow\rangle)\nonumber\\
&|Q_{-\frac{3}{2}}\rangle=|\downarrow\downarrow\downarrow\rangle\nonumber\\
&|\Delta_{+\frac{1}{2}}\rangle=\frac{1}{\sqrt{4\Omega^2+2\Omega(J'-2J_2+J_1)}}\left\{(J_2-J'-\Omega)|\uparrow\uparrow\downarrow\rangle+(J_1-J_2+\Omega)|\uparrow\downarrow\uparrow\rangle+(J'-J_1)|\downarrow\uparrow\uparrow\rangle\right\}\nonumber\\
&|\Delta_{-\frac{1}{2}}\rangle=\frac{1}{\sqrt{4\Omega^2+2\Omega(J_2-2J'+J_1)}}\left\{(J_2-J'+\Omega)|\uparrow\downarrow\downarrow\rangle+(J'-J_1-\Omega)|\downarrow\uparrow\downarrow\rangle+(J_1-J_2)|\downarrow\downarrow\uparrow\rangle\right\}\nonumber\\
&|\Delta'_{+\frac{1}{2}}\rangle=\frac{1}{\sqrt{4\Omega^2+2\Omega(2J_2-J'-J_1)}}\left\{(J_2-J'+\Omega)|\uparrow\uparrow\downarrow\rangle+(J_1-J_2-\Omega)|\uparrow\downarrow\uparrow\rangle+(J'-J_1)|\downarrow\uparrow\uparrow\rangle\right\}\nonumber\\
&|\Delta'_{-\frac{1}{2}}\rangle=\frac{1}{\sqrt{4\Omega^2+2\Omega(2J'-J_2-J_1)}}\left\{(J_2-J_1-\Omega)|\uparrow\downarrow\downarrow\rangle+(J'-J_1+\Omega)|\downarrow\uparrow\downarrow\rangle+(J_1-J_2)|\downarrow\downarrow\uparrow\rangle\right\},
\end{align}
\end{widetext}
the relative energies are given by
\begin{align}\label{eigl}
&E_{Q_{S_z}}=\frac{J'+J_1+J_2}{4}\nonumber\\
&E_{\Delta_{S_z}}=-\frac{J'+J_1+J_2-\Omega}{2}\nonumber\\
&E_{\Delta'_{S_z}}=-\frac{J'+J_1+J_2+\Omega}{2},
\end{align}
where $\Omega=\sqrt{J'^2+J_2^2+J_1^2-J'J_2-J_1J_2-J'J_1}$.

At this point we focus our attention on some cases of physical interest. The cases under study correspond to two different situations: one in which the coupling $J_2$ is greater with respect to the others, which means that two electrons are confined in the right dot, and the opposite situation in which two electrons are in the left dot, that means a larger contribution from $J'$. The explicit expressions of eigenvectors and eigenvalues in such limiting cases are collected in Appendix \ref{B}.

\begin{figure}[h]
\begin{center}
\includegraphics[width=0.3\textwidth]{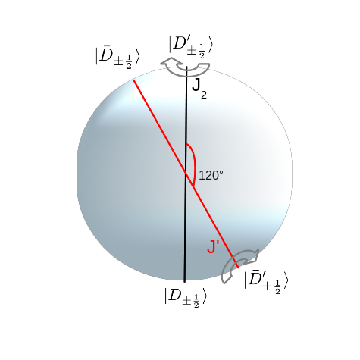}
\end{center}

\caption{Graphical representation of the eigenvectors on the Bloch Sphere}\label{bloch}
\end{figure}

States of the qubit correspond to points on the Bloch sphere shown in Fig. \ref{bloch}. Exchange $J_2$ drives rotations about the vertical axis, where $|D'_{S_z}\rangle$ and $|D_{S_z}\rangle$, given in Eqs.(\ref{b1}), denote respectively states with the pair of spins in the right dot forming a singlet and a triplet states. On the other hand exchange $J'$ drives rotations about an axis tilted by 120° and defined by doublets $|\bar{D}'_{S_z}\rangle$ and $|\bar{D}_{S_z}\rangle$, see Eqs.(\ref{b3}), that correspond to singlets and triplets states on the left dot. Arbitrary single-qubit operations can be achieved by concatenating up to four exchange pulses as shown in Ref.\cite{spin2}.

The logical states introduced in Eq.(\ref{01}) are equivalent to
\begin{equation}
|0\rangle\equiv|\bar{D}'_{-\frac{1}{2}}\rangle, \qquad |1\rangle\equiv|\bar{D}_{-\frac{1}{2}}\rangle
\end{equation}
in the Bloch sphere. The logical basis is formed from two states with equal $S_z$, one taken from each doublet. The implementation of full single qubit operation needs the preparation and read out of only two of the three electron spins. 

We point out that the system under study and the linear triple dot architecture proposed in Ref.\cite{spin2}, due to the fact that contain three interacting electrons, have the same symmetries, providing an analogue picture in the Bloch sphere. This means that it is possible to define gate operations like in the triple dot model. However due to the different architectures, the Hamiltonian models present some distinctions. First of all in the triple dot it is not allowed to spins in the external dots to interact directly, as their interaction is indeed mediated by the spin in the middle. In addition the parameters involved in the double dot are influenced by the confinement of two electrons in the same dot, as occurs for $t_{12}=0$. These differences determine a change in the expressions for eigenvalues and eigenvactors, and leave unchanged the shape. The convenience in adopting the double dot model with respect to the triple dot rely on a great advantage from a practical point of view, requiring the control of two dots instead of three.
 
\section{Conclusions}
Starting from the generalized Hubbard model, we analytically derive in the Schrieffer-Wolff transformation framework a complete formula for the exchange coupling between three confined electrons in a double quantum dot, independently of the host material. The tunable exchange couplings depend explicitly on intradot and interdot Coulomb repulsions, tunnel couplings and interdot bias. The three interacting electrons are used to encode a logical qubit, that combine singlet-triplet states of the pair of spins in the left dot with single spin states of the electron spin in the right dot.

We have shown how the exact dynamics of the system is exploited to guide the system towards specific states of interest, by suitably tuning the external bias voltages and the tunneling parameters. Moreover the effective Hamiltonian is exploited to describe a realistic silicon qubit studying the stationary solutions. We have also provided a picture of the system on the Bloch sphere to link the tuning of the effective exchange coupling with the rotation of the involved state. 

Our results pave the way towards two-qubit operations controlled by the effective exchange couplings, in order to implement complex logic gates.

\acknowledgements
This work is partially supported by the project QuDec, granted by the Ministero della Difesa of Italy. 

\appendix
\section{Dynamical evolution}\label{A}
Time-dependent Schr\"odinger equation for the hybrid qubit described by Hamiltonian (\ref{effmatrix}) is here solved. 

The state of the system at the initial time $t=0$ is written as a normalized superposition of the states of the logical basis $\{|0\rangle,|1\rangle\}$ with probability amplitudes given by $a(0)$ and $b(0)$. The normalization condition $|a(0)|^2+|b(0)|^2=1$ is satisfied. Due to the conservation of the total angular momentum operator, it follows that also at a generic time instant $t$, the state of the system can be written analogously with probability amplitudes $a(t)$ and $b(t)$ depending explicitly on time
\begin{equation}
|\psi(0)\rangle=a(0)|0\rangle+b(0)|1\rangle\quad\Rightarrow|\psi(t)\rangle=a(t)|0\rangle+b(t)|1\rangle.
\end{equation}
By inserting this expression into the time-dependent Schr\"odinger equation $H|\psi(t)\rangle=i|\dot{\psi}(t)\rangle$ and by solving the system of two first order differential equations for the probability amplitudes $a(t)$ and $b(t)$, we finally obtain
\begin{equation}\label{ab}
  \left\{
  \begin{array}{l l}
    a(t)=c_1e^{\lambda_1t}+c_2e^{\lambda_2t}\\
    b(t)=c_1\frac{\lambda_1-iA}{iC}e^{\lambda_1t}+c_2\frac{\lambda_2-iA}{iC}e^{\lambda_2t},\\
  \end{array} \right.
\end{equation}
where
\begin{equation}\label{coeff}
\lambda_{1,2}=i(\alpha\mp\beta), \quad \alpha=\frac{A+B}{2}, \quad \beta=\frac{\sqrt{(A-B)^2+4C^2}}{2}
\end{equation}
and
\begin{equation}\label{abc}
A=\frac{3}{4}J', \quad B=\frac{\sqrt{3}}{4}(J_1-J_2), \quad C=-\frac{1}{4}J'+\frac{1}{2}(J_1+J_2).
\end{equation}
Eqs.(\ref{ab}) contains the more general form for the probability amplitudes at every time instant $t$. Once that the initial condition is fixed it is possible to extract the values for the coefficients $c_1$ and $c_2$.

In the case of the specific initial condition analyzed in Section II in which the system is prepared in the state of the logical basis corresponding to $|\psi(0)\rangle=|0\rangle$, the coefficients are
\begin{equation}
  \left\{
  \begin{array}{l l}
    a(0)=1\\
    b(0)=0.\\
  \end{array} \right.
\end{equation}
After straightforward calculations we get the probability amplitudes
\begin{equation}\label{10}
  \left\{
  \begin{array}{l l}
    a(t)=\frac{e^{i\alpha t}}{\beta}\left[\beta\cos(\beta t)+i(A-\alpha)\sin(\beta t)\right]\\
    b(t)=-ie^{i\alpha t}\frac{(A-\alpha)^2-\beta^2}{\beta C}\sin(\beta t).\\
  \end{array} \right.
\end{equation}

\section{Eigenvalues and eigenvectors of three exchange-coupled spins in two limiting cases of interest}\label{B}
In this Appendix eigenvectors and eigenvalues of the hybrid qubit, decribed by the effective Hamiltonian (\ref{Heff}), are presented in two special cases. Two limiting conditions of interest from the practical point of view, are analyzed.
\begin{enumerate}
\item{Case $J_2\gg J'\simeq J_1$}

Under the condition on the exchange coupling $J_2\gg J'\simeq J_1$, that means that two electron are confined in the right dot, eigenvectors and eigenvalues in Eqs.(\ref{eig}) and (\ref{eigl}) become
\begin{align}\label{b1}
&|D_{+\frac{1}{2}}\rangle=\frac{1}{\sqrt{6}}\left(|\uparrow\uparrow\downarrow\rangle+|\uparrow\downarrow\uparrow\rangle-2|\downarrow\uparrow\uparrow\rangle\right)\nonumber\\
&|D_{-\frac{1}{2}}\rangle=\frac{1}{\sqrt{6}}\left(|\downarrow\downarrow\uparrow\rangle+|\downarrow\uparrow\downarrow\rangle-2|\uparrow\downarrow\downarrow\rangle\right)\nonumber\\
&|D'_{+\frac{1}{2}}\rangle=\frac{1}{\sqrt{2}}\left(|\uparrow\uparrow\downarrow\rangle-|\uparrow\downarrow\uparrow\rangle\right)\nonumber\\
&|D'_{-\frac{1}{2}}\rangle=\frac{1}{\sqrt{2}}\left(|\downarrow\downarrow\uparrow\rangle-|\downarrow\uparrow\downarrow\rangle\right)
\end{align}
\begin{align}
&E_{D_{S_z}}=\frac{1}{4}J_2\nonumber\\
&E_{D'_{S_z}}=-\frac{3}{4}J_2.
\end{align}
\item{Case $J'\gg J_2\simeq J_1$}

On the other hand, the opposite condition corresponding to two electrons confined in the left dot, that is $J'\gg J_2\simeq J_1$, gives as eigenvectors and eigenvalues 
\begin{align}\label{b3}
&|\bar{D}_{+\frac{1}{2}}\rangle=\frac{1}{\sqrt{6}}\left(|\downarrow\uparrow\uparrow\rangle+|\uparrow\downarrow\uparrow\rangle-2|\uparrow\uparrow\downarrow\rangle\right)\nonumber\\
&|\bar{D}_{-\frac{1}{2}}\rangle=\frac{1}{\sqrt{6}}\left(|\uparrow\downarrow\downarrow\rangle+|\downarrow\uparrow\downarrow\rangle-2|\downarrow\downarrow\uparrow\rangle\right)\nonumber\\
&|\bar{D}'_{+\frac{1}{2}}\rangle=\frac{1}{\sqrt{2}}\left(|\uparrow\downarrow\uparrow\rangle-|\downarrow\uparrow\uparrow\rangle\right)\nonumber\\
&|\bar{D}'_{-\frac{1}{2}}\rangle=\frac{1}{\sqrt{2}}\left(|\downarrow\uparrow\downarrow\rangle-|\uparrow\downarrow\downarrow\rangle\right)
\end{align}
\begin{align}
&E_{\bar{D}_{S_z}}=\frac{1}{4}J'\nonumber\\
&E_{\bar{D}'_{S_z}}=-\frac{3}{4}J'.
\end{align}
\end{enumerate}

\end{document}